\documentclass[runningheads]{llncs}
\usepackage{multirow}
\usepackage{graphicx}
\usepackage{amsmath}
\usepackage[colorlinks,citecolor=red,urlcolor=blue,bookmarks=false,hypertexnames=true]{hyperref}     

\begin{document}
\title{MIST GAN: Modality Imputation using Style Transfer for MRI}

\titlerunning{MIST GAN}

\author{Jaya Chandra Raju\inst{1,2}\and
Kompella Subha Gayatri\inst{1,2}\and
Keerthi Ram\inst{2}\and
Rajeswaran Rangasami\inst{3}\and
Rajoo Ramachandran\inst{3}\and
Mohanasankar Sivaprakasam\inst{1,2}
}

\authorrunning{J.C.Raju et al.}

\institute{Indian Institute of Technology Madras\and
Healthcare Technology Innovation Centre\and
Sri Rama Chandra Institute of Higher Education and Research
}
\maketitle              

\begin{abstract}
MRI entails a great amount of cost, time and effort for generation of all the modalities that are recommended for efficient diagnosis and treatment planning. Recent advancements in deep learning research show that generative models have achieved substantial improvement in the aspects of style transfer and image synthesis. In this work, we formulate generating the missing MR modality from existing MR modalities as an imputation problem using style transfer. With a multiple-to-one mapping, we model a network that accommodates domain specific styles in generating the target image. We analyse the style diversity both within and across MR modalities. Our model is tested on the BraTS'18 dataset and the results obtained are observed to be on par with the state-of-the-art in terms of visual metrics, SSIM and PSNR. After being evaluated by two expert radiologists, we show that our model is efficient, extendable, and suitable for clinical applications.
\end{abstract}

\section{Introduction}
MRI is a versatile imaging technique that can image in multiple sequences each highlighting specific tissue type based on the TR (Repetition time) and TE (Time to Echo) parameters. The most commonly generated sequences are T1 (with optional gadolinium enhancement), T2 and FLAIR (Fluid-Attenuated Inversion Recovery), which are primarily used for diagnostic purposes in a clinical setting. It is stated in\cite{Drevelegas2011Imaging} that in brain MRI, MR images with T1, T2, or FLAIR contrast, are all required for accurate diagnosis and segmentation of cancer margin. At the same time, acquiring all the sequences from a patient is also an extremely tedious task, especially given the long scanning time and the added cost. Even if all the sequences are acquired, there could be faulty scans with distorted contrast which often are a result of motion artifacts or noise. T1c requires injecting a dye which may induce side effects and is not recommended for pregnant women or patients with kidney or heart issues. Considering these clinical limitations and feasibility, it is therefore necessary to explore the potential of cross modality synthesis for efficient data infilling with respect to MRI.

Generating a new MRI modality from the existing modalities can be formulated as an image-to-image translation problem, whose goal is to learn a mapping between two different visual domains\cite{isola2018imagetoimage}. Here, domain implies a set of images that can be grouped as a visually distinctive category and each image has a unique appearance which is the style. For example, in MR images, each sequence can be considered as a domain and different variants in each sequence can be considered as different styles. In recent years, Generative Adversarial Networks (GANs)\cite{goodfellow2014generative} have been extensively studied for image-to-image translation applications and showed promising outcomes\cite{cvpr2018_stargan}\cite{zhu2020unpaired-cycleGAN}\cite{kim2017learning-DiscoGAN}. Lately, cross domain image translation using GANs have been considered for MRI. Due to limited scalability, i.e. requiring N(N-1) generators for N-domain image transfer, many networks have been restricted for one-to-one synthesis\cite{Yang_MRI_Cross_Modality}\cite{Dar_multi_contrast}. The next formulation of significance is multiple input-to-multiple target synthesis of which models such as MM-GAN\cite{tmi2020_mmgan}, AutoSyncoder\cite{Raju2020AutoSyncoder}, latent representation\cite{tmi2018_latentrepr},have performed efficiently, but these networks remain either bulky or in need for a target based training. However, in medical image synthesis, multiple-to-one mapping models are more relevant as learning proprietary information of individual modalities can be more synergistic\cite{mi2020_collaGAN}\cite{li_diamondgan_2019}. They affirm that a more accurate target image is generated when information from all the input modalities is used for training.

Though the aforementioned recent works successfully addressed the scalability issue for multi-domain MRI synthesis/imputation, none of the models consider the style diversity within the domain. Various MRI imaging centres follow various imaging protocols giving multiple possible combinations of sequence parameters each of which can be considered as a different style. Learning these styles and being able to generate a target image of the required style would greatly improve the comparability between two scans. For example, in multiple sclerosis (MS) patients, treatment planning and decisions are majorly based on longitudinal comparisons of MRI studies. Also, existing lesion quantification tools require completely identical modalities scanned at multiple time points\cite{li_diamondgan_2019}. It is therefore important to extract and reflect this style information. For this, a very recent framework, StarGAN v2\cite{choi2020starganv2} introduced the concept of style code to represent diverse styles of a specific domain and it is learnt using two additional modules, style encoder and mapping network. Another recent work ReMIC\cite{ReMIC}, used style based representation disentanglement for image completion and segmentation in MRI.

\noindent\textit{Contribution:} 1) We propose MIST GAN for MRI, which is a unified domain scalable architecture that learns the multiple-to-one cross-modality mapping with just a single generator and discriminator. 2) We establish that our model is capable of learning and reflecting the diverse styles present within a domain. We also prove that, contrary to the popular assumption, an MRI modality is not just a single style, but a spectrum of styles.

\begin{figure}[t]
\begin{center}
\includegraphics[width=90mm, scale=0.25]{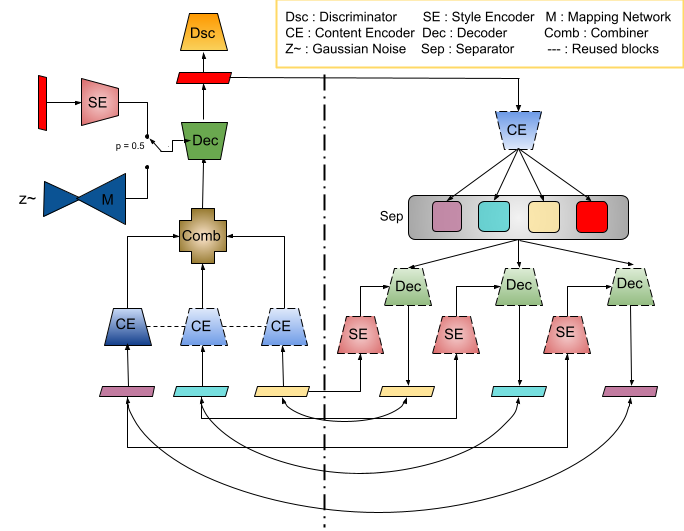}
\caption{MIST GAN overview. Left: Generates a target image from the given three input images which is of the style specified by \textit{SE}. Right: The cycle to reconstruct the input images from the generated image with the help of \textit{Sep} and reusable \textit{CE}, \textit{SE} and \textit{Dec}.}
\label{fig:architecture}
\end{center}
\end{figure}

\section{Proposed Framework}
Let $D = \{p,q,r,..\}$ be the set of  domains and $x_d$ be the image drawn from domain $d \in D$. Given the target domain, $t$ that is randomly sampled from $D$, the set of input domains will be  $I = D - \{t\}$.  When set of input images, $X_d = \{x_d|d \in I \}$ and a target image $x_t$ is given, the goal of the network is to learn the mappings from multiple input modalities to the target modality while also learning varied styles within the domain. The following modules are included in our network to realize the same:

\noindent\textbf{Generator (G):}

\textit{Style encoder (SE):} extracts the fine style features present in a given reference image. $SE$ takes the image, $x_d$ and the domain information, $d$, as input to generate the domain specific style code $s_d = SE(x_d,d)$.

\textit{Mapping Network (M):} is a fully connected network that learns to map random Gaussian noise ($z$), to latent style code distribution, $s_d = M(z,d)$. In applications where the target image is not given for the style encoding, a random target domain specific style code can be generated by using this mapping network which is guided by the discriminator($Dsc$) during the training.

\textit{Content Encoder (CE):} is expected to take in an image, $X$, and generate its content, $c = CE(x)$, in latent representation which is modality invariant. Instance normalization (IN) is used for the down-sampling blocks.

\textit{Decoder (Dec):}
Given the content($c$) generated by $CE$ our decoder tries to generate the target image, $\hat{x}_t = Dec(c,s_t)$, reflecting the provided target style code $s_t$. Here, we use Adaptive Instance Normalization(AdaIN)\cite{huang2017arbitrary} for the up-sampling blocks and add the skip connections through a high-pass filter.

\textit{Combiner and Separator (Comb, Sep):} The combiner generates a combined content, $cc = Comb(concatenate(c_p, c_q, c_r))$ by integrating the content generated by $CE$ for different domain images, $c_p, c_q, c_r$ where $p,q,r \in I$. \textit{cc} stores the useful key features extracted from all the input modalities. The target image $\hat{x}_t$ is generated by passing combined content, $cc$, through the $Dec$. Once we have generated the target image, $\hat{x}_t$, it is encoded again using the $CE$ in order to get the new combined content, $\hat{cc}$. Now, the separator takes in this content ($\hat{cc}$) and the modality information ($d$) of each input and generates domain specific content, $c_d = Sep(\hat{cc}, d)$. This is given to the $Dec$ that generates back the original input images (Fig. \ref{fig:architecture}).

\noindent\textbf{Discriminator (Dsc)}
used in this network is a multi-task discriminator with multiple output branches where each branch learns a binary classification determining whether an image is a real image, ($x_d$), or a decoder outputted fake image, ($\hat{x}_t$). It takes in an image($x_d$) and its domain information($d$), and predicts the probability($p_{r/f}$)$= Dsc(x_d/\hat{x}_t; d/t)$ of how real the given image is.
\subsubsection{Training Objectives:}
\paragraph{Cyclic loss:}
To ensure that the network properly preserves the domain invariant characteristics of the given inputs, we enforce the Dec to reconstruct the input images from the target image by incorporating a cyclic loss ($CL = CSL+CCL$) which consists of  cyclic style loss($CSL$) and cyclic content loss($CCL$). $CSL$ is the loss defined between the style information encoded from the generated fake image and the style information encoded from the target image sampled.
To synthesize the input images from the imputed target image, $\hat{x}_t$, $CE$, $Sep$ are used to get the corresponding modality specific content and $SE$, $Dec$ are reused to generate the respective images, $\hat{x}_p$, $\hat{x}_q$ and $\hat{x}_r$. Now, $CCL$ is defined as the summation of domain-wise $L1$ loss between translated images and source images.
\begin{equation}
    CSL = ||SE(x_t) -SE(\hat{x}_t)||_1
\end{equation}
 \begin{equation}
     CCL =\sum_{d = p,q,r} ||\hat{x}_d-x_d||_1  
 \end{equation}
\paragraph{Adversarial loss:}
The adversarial loss is defined to impose the network to produce images that are indistinguishable from the true distribution. The adversarial loss components are different for the generator part ($G_{adv}$) and the $Dsc$ ($Dsc_{adv}$) as shown below:
\begin{equation}
     G_{adv} = 1 - Dsc(\hat{x}_t, t)
\end{equation}
 \begin{equation}
    Dsc_{adv} = (1 - Dsc(x_t, t)) + (0 - Dsc(\hat{x}_t, t))
 \end{equation}
\paragraph{Style Diversification Loss (SDL):}
To enable the network to produce diverse styled images, we regularize it with the diversity sensitive loss\cite{choi2020starganv2}. Two reference images are randomly sampled from the target domain ($x_{t1};x_{t2}$) to get $s_{t1}; s_{t2}$  estimated by $SE$. We generate images using these two styles and maximize the difference between them which forces the network to learn meaningful style features. The total objective can be summarized for generator and discriminator as follows:
\begin{equation}
\boldsymbol{\min_{G}} (CL + G_{adv} - SDL) \quad \textrm{and}\quad \boldsymbol{\min_{Dsc}}(Dsc_{adv})
\end{equation}

\begin{figure}[t]
\begin{center}
\includegraphics[width=0.9\textwidth, scale=0.25]{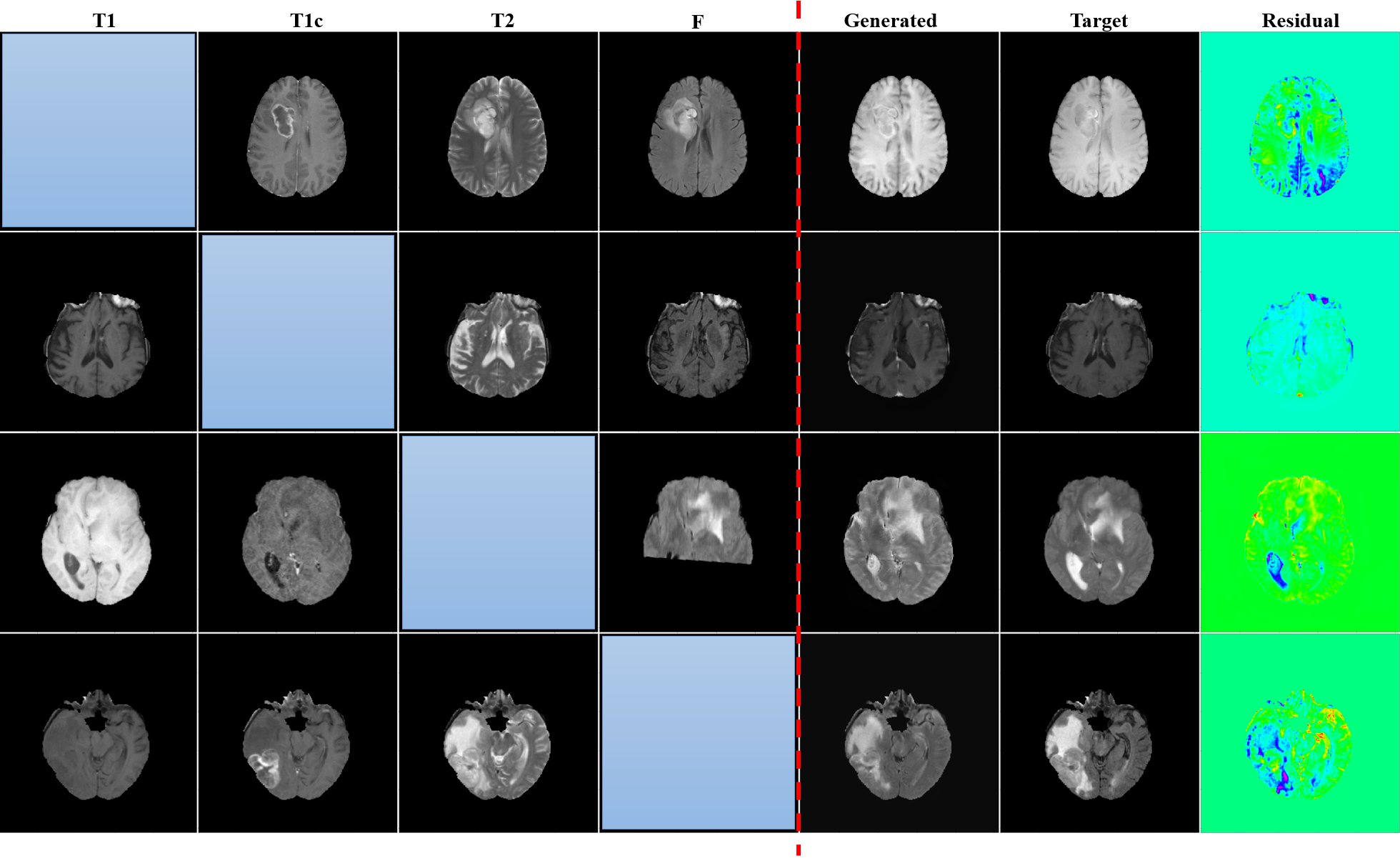}
\caption{Qualitative results achieved by the proposed model on the HGG BraTS'18 dataset. A sample of the domain-wise generated images along with their respective ground truths are shown. The blue squares represent the missing/target modality.}
\label{fig:imputation_results}
\end{center}
\end{figure}
\begin{center}
\begin{table*}[hbt!]
\caption{Quantitative performance of the model on BraTS'18 HGG and LGG cohort.}
\centering
\resizebox{\textwidth}{!}{\begin{tabular}{|c|c|c|c|c|c|}
\hline
 & Modality & Metrics & Stargan-v2 & MM-GAN & Proposed\\ \hline 
\multirow{8}{*}{LGG} & \multirow{2}{*}{T1} & SSIM & 0.90143 $\pm $ 0.01685& \textbf{0.9288 $\pm$ 0.0137}
& 0.90193 $\pm$ 0.02157\\ \cline{3-6} & 
& PSNR & 22.96876 $\pm$ 3.28355 & \textbf{25.9841 $\pm$ 2.1926}
& 24.68469 $\pm$ 3.87952\\ \cline{2-6} &
\multirow{2}{*}{T1c} & SSIM & 0.89411 $\pm $ 0.01774 & \textbf{0.9086 $\pm$ 0.0253} & 0.89597 $\pm$ 0.02294\\ \cline{3-6}& 
& PSNR & 24.61375 $\pm$ 2.66390 & 22.4980 $\pm$ 1.1684 
& \textbf{25.07036 $\pm$ 3.43035}\\ \cline{2-6} &
\multirow{2}{*}{T2} & SSIM & 0.88846 $\pm $ 0.02120 & \textbf{0.9030 $\pm$ 0.0303} & 0.88411 $\pm$ 0.02094\\ \cline{3-6} & 
& PSNR & 24.60764 $\pm$ 2.88763 & \textbf{25.6005 $\pm$ 2.7909}
& 24.761391 $\pm$ 2.57905\\ \cline{2-6} &
\multirow{2}{*}{F} & SSIM & 0.86251 $\pm $ 0.02914 & 0.8692 $\pm$ 0.0224
& \textbf{0.87498 $\pm$ 0.02930}\\ \cline{3-6} & 
& PSNR & 22.67723 $\pm$ 2.65003 & 23.0852 $\pm$ 1.5142
& \textbf{ 23.72480 $\pm$ 2.86680}\\ \hline 
\multirow{8}{*}{HGG} & \multirow{2}{*}{T1} & SSIM
& 0.88832 $\pm $ 0.01787& \textbf{0.9228 $\pm$ 0.0190}
& 0.88226 $\pm$ 0.02094\\ \cline{3-6} & 
& PSNR & 23.99222 $\pm$ 2.58524 & 24.173 $\pm$ 3.2754
& \textbf{25.62511 $\pm$ 1.94889}\\ \cline{2-6} &
\multirow{2}{*}{T1c} & SSIM & 0.87734 $\pm $ 0.02066 & \textbf{0.9239 $\pm$ 0.0375} & 0.87303 $\pm$ 0.02411\\ \cline{3-6}& 
& PSNR & 24.09959 $\pm$ 3.23823 & 24.372 $\pm$ 2.2792 
& \textbf{26.42274 $\pm$ 1.80974}\\ \cline{2-6} &
\multirow{2}{*}{T2} & SSIM & 0.86777 $\pm $ 0.01812 & \textbf{0.9349 $\pm$ 0.0262} & 0.87160 $\pm$ 0.02162\\ \cline{3-6} & 
& PSNR & 23.45378 $\pm$ 2.26071 & \textbf{28.678 $\pm$ 2.3290}
& 24.88888 $\pm$ 1.33794\\ \cline{2-6} &
\multirow{2}{*}{F} & SSIM & 0.84534 $\pm $ 0.02543 & \textbf{0.9150 $\pm$ 0.0275}
& 0.85406 $\pm$ 0.02861\\ \cline{3-6} & 
& PSNR & 21.95010 $\pm$ 2.74631 & \textbf{26.397 $\pm$ 1.9733}
&  24.79925 $\pm$ 2.10912\\ \hline 
\end{tabular}}
\label{tab:imputation_metrics}
\end{table*}
\end{center}
\begin{figure}[t]
\begin{center}
\includegraphics[width=120mm, scale=0.25]{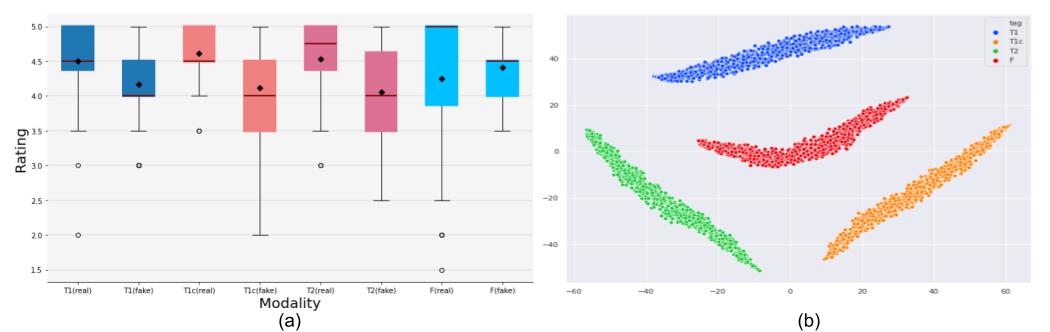}
\caption{(a):Validation scores assigned by expert radiologists for all the modalities when blinded to the ground truth label of the image. (b):t-SNE plot showing disengagement of style codes across four domains and a spectrum of styles within each domain.}
\label{fig:rating_and_tsne}
\end{center}
\end{figure}
\begin{figure} 
\begin{center}
\includegraphics[width=0.9\textwidth, scale=0.25]{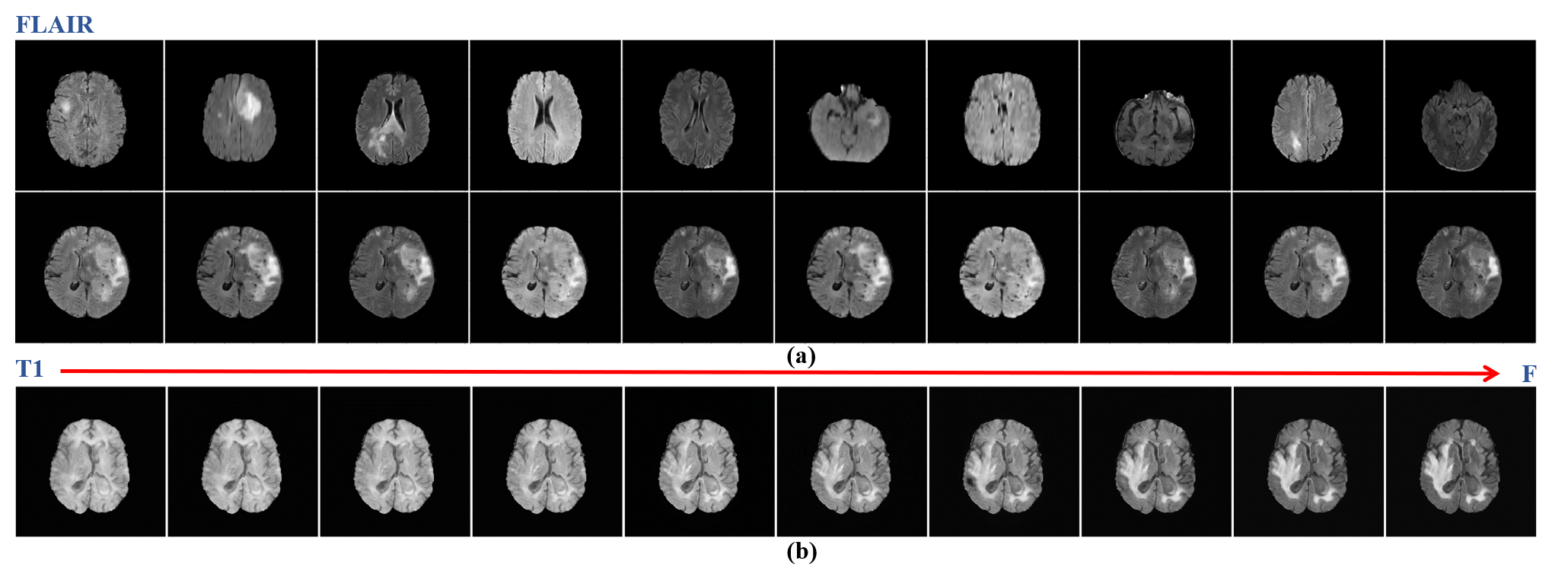}
\caption{(a) Representation of style forms: the second row shows the Flair generated images when its corresponding first row image is given as a reference image to the style encoder. (b) Style interpolation: T1 to F transition, achieved by linearly interpolating the styles between mean style of T1 and mean style of Flair in steps of 0.1.}
\label{fig:style_demo}
\end{center}
\end{figure}

\section{Results and Discussion}

\textbf{Dataset and Implementation:} The performance of the proposed architecture is evaluated on BraTS'18 dataset\cite{menze:hal-00935640}\cite{MenzeBjoernHetal} that consists of 75 LGG and 210 HGG patient records of T1, T1c, T2 and Flair modalities all of which are axially acquired, co-registered and skull stripped. The dataset is split for training, validation and testing in the ratio 3:1:1. All the slices are padded to a uniform size of 256 x 256 and the intensity values are linearly normalized between (0,1). For the training, the batch size is set to 2 and the model is trained for 200k iterations. The training duration is approximately two days on a single NVIDIA GEFORCE GTX 1080 Ti GPU, occupying about 10GB. Adam\cite{Adam} optimizer is adopted for training and the learning rate is set to 1e-4 for all modules except mapping network for which it is set as 1e-6. All weights are initialized using \cite{He_Initialization}.

\noindent\textbf{Qualitative and quantitative analysis:} We generate a target modality that not only learns relevant information from multiple input modalities, but also reflects the user specified style. Any 3 of 4 MRI modalities, irrespective of their ordering are taken as inputs and the missing modality is imputed using style transfer. We ensure domain scalability and limit the network complexity by reusing the core blocks ($CE$, $SE$, $Dec$). The model is studied for both HGG and LGG datasets of BraTS’18 and quantitatively evaluated using PSNR (Peak Signal-to-Noise Ratio) and SSIM (Structural Similarity Index) metrics. The results for the same are shown in Table \ref{tab:imputation_metrics}. Based on these metrics it is observed that our synthesized T1c is closer to the real T1c for both HGG and LGG, thereby opening the possibility of generating a contrast enhanced image without actually injecting the contrast agent, an application which is of high clinical relevance. 

We compare the achieved results with the results of a baseline model, MM-GAN, which incorporates a target based training with supervised loss. Despite being blind to the target image and solely guided by the discriminator, our model performed on par, better for few modalities (LGG - T1c, F; HGG - T1, T1c) in terms of PSNR (Table \ref{tab:imputation_metrics}). We have also implemented StarGAN v2, a style based model for MRI and recorded the metrics for comparison. We trained the model as is, but while testing, instead of just giving one image, we reused the encoder of the generator for multiple inputs and averaged the content in the latent space and passed it to the decoder to get the generated image. It can be seen from Table \ref{tab:imputation_metrics} that the proposed model clearly outperformed StarGAN v2. A sample of our modality-wise generated output is shown in Fig. \ref{fig:imputation_results} and it can be observed that, generated T1 and T1c are enhanced and more sharp than their respective ground truths, a proper T2 is generated even though one of its source images is faulty and generated Flair is quite close to its original image.

\noindent\textbf{Style Analysis:} In order to study the efficiency of our model in learning, extracting and reflecting varied styles, we attempted generating different style forms of a single image in a given domain. We fed the network with any arbitrary style image and made it generate the same target image in the specified random style. Fig. \ref{fig:style_demo}(a) shows the style forms for Flair domain: top row shows the given reference style images and the bottom row shows the corresponding generated target images. Clear variations can be noticed in the generated images indicating that any modality is not just a single style but a spectrum of styles and our model succeeded in capturing them. The same is evident from the t-SNE(t-distributed stochastic neighbor embedding) plot Fig. \ref{fig:rating_and_tsne}(b) of the latent style codes, in which it is also seen that style information is properly disentangled between the four domains. With the acquired domain specific mean style value, we observed the transition of a generated image between two domains, while gradually changing the style (simple weighted average/linear interpolation of latent style), thereby enabling operator, the flexibility to choose and change between the styles. Fig. \ref{fig:style_demo}(b) shows one such smooth transition achieved between T1 and Flair. Indeed an unseen intermediate modality which is neither completely T1 or Flair has surfaced in this interpolation whose clinical utility could be a prospective research. However, some droplet artifacts are noticed when interpolated with intermediate styles which is probably due to a harsh style diversification loss. For quantitative analysis, mean FID (Frechét inception distance)\cite{heusel2018gans} and mean LPIPS (Learned perceptual image patch similarity)\cite{zhang2018perceptual} are extracted for all modalities. The mean FID  (between the generated images and the true images) and LPIPS  (across the generated images) scores for T1, T1c, T2 and Flair are [49.37, 70.52, 49.66, 74.25] and [0.0159, 0.0261, 0.0092,  0.0277] respectively when assessed for BraTS'18 dataset.

\noindent\textbf{Visual Evaluations by Radiologists:} 
For qualitative assessment of the achi- -eved results, we consulted two expert radiologists with 10+ years of professional experience. They each evaluated the output image on a scale of 5 (1, being very poor and 5, being excellent). In each trial, they were provided with four images, the left three were the source images and on the right, a corresponding target image was shown, which is either a real image or a generated one. Radiologists are blinded to the nature (real or fake) of the image so as to avoid any bias and to collect a more objective feedback. Each of them are presented with 400 such random trials for all the four modalities. The rating was given based on the image quality, capturing of small sub-structures and efficiency in mapping the source images to the target image. The ratings of the two radiologists are averaged and the boxplot of the results are shown in Fig. \ref{fig:rating_and_tsne}(a). From the plot it is noticed that for T1, T1c and T2, the mean rating of synthesized images lags the mean rating of real images by about 0.5. But for the Flair modality, the generated images were given a better rating when compared to the real images. Both the experts acknowledged that the synthetic Flair images are as good as the original flair. The radiologists’ opinion was also taken on the performed style analysis of the output images and they both affirmed that the model captured multiple styles within each domain and that such work would help in longitudinal MR comparability studies of patients with progressive neurological conditions.
\section{Conclusion}
Motivated by the need to create an efficient and novel method for MRI modality synthesis, we have derived a style transfer formulation for imputation and realized an efficient GAN based network that generates an enhanced version of the missing MR modality from given three input modalities. We show that our network is not only scalable to any number of input modalities, but also capable of picking up style variations within each modality. Evaluation on the BraTS'18 multimodal brain MRI dataset suggests that the method is promising and opens new avenues for further research. 

\bibliographystyle{splncs04}
\bibliography{my_bibliography}
\end{document}